\documentclass[12pt,preprint]{aastex}
\shorttitle{Are GRBs the most violent events?}
\shortauthors{L\'{o}pez E. D.}

\begin{document}

\title{Are GRBs the most violent events in the Universe?}

\author{Ericson D. L\'{o}pez}
\affil{Harvard Smithsonian Center for Astrophysics \\
60 Garden Street, Cambridge, MA 02138, USA \\
and\\
Observatorio Astron\'{o}mico de Quito \\ 
Escuela Polit\'{e}cnica Nacional\\
Interior del Parque La Alameda, \\
Av. Gran Colombia s/n, 17-01-165, Quito - Ecuador}

\begin{abstract}
Considering a GRB event as a relativistic flux where the relativistic beam makes radiation become anisotropic, we are able to show that the required intrinsic energy associated with these events is significantly smaller than those values commonly presented in literature. Our results show energy values around $10^{44}$ ergs for Lorentz $\Gamma$  factor $\sim 10$ and around $10^{38}$ ergs for $\Gamma \sim 300$, values which are more compatible with energies involved in AGN events rather than those related to the formation of stellar black holes and hypernovas.
\end{abstract}


\section{Introduction}

The BATSE detector observations on the Compton Gamma-Ray Observatory have been conducting our understanding about the origin of GRBs to be cosmologic. It proved that they are not confined to the Galactic plane, but they are instead isotropically distributed. Actually, it is well accepted the extragalactic origin of the GRB emitters which produce bursts with energies about $10^{51}$ - $10^{53}$ ergs in few seconds of $100 $KeV- $1 $MeV photons.

Several difficulties have been found in this context, being one of the most important the problem of  high opacity for all photons about the $\gamma$ - $\gamma$ pair creation threshold (see Fireball model by  \citet{pi93} ).  For cosmological GRBs an instantaneous energy of $10^{51}$ ergs is implied from the observed flux of about $10^{-7}$  ${ergs\over {cm^2 sec}}$, for this value both the absorptive and scattering optical depths are very large and it is extremely difficult to understand how photons about the   pair creation threshold escape from the emitting region located close to the compact source.
The optical thickness problem initially was associated only with cosmological models of GRBs, but later it was showed that it is also a problem present in Galactic GRBs of a constant luminosity  $L \sim 10^{41}$ ${ergs\over {sec}}$ (see Piran \& Shemi, 1993).

Several authors have proposed to follow the only consistent and available  way in order to solve this fundamental problem. This means, analyzing the problem of large opacity as a relativistic illusion causes by the emitting plasma of $\gamma$-rays which is moving relativistically as a whole \citep{de89,ho89,kr91}.  The models involving the relativistic bulk motion of a
$\gamma$-ray emitting plasma gives us a plausible interpretation in order to match the theory with high-energy photons observed to escape in GRB events. The results presented in the  \citet{kr91} paper are remarkable since it is clearly exposed the   relativistic bulk motion potential of an emitting plasma for providing an elegant solution to the problem of large opacity.

In the current work, we follow this physical alternative working-out a GRB model by taking into consideration a relativistic moving plasma.  Then, the main parameters, which characterize the physical conditions of the emitting material, must be reduced or boosted by a suitable potency of the Doppler Lorentz factor  $\delta = {1/ {\gamma (1-\beta \mu)}}$, where $\gamma = {1\over{\sqrt{1- \beta ^2}}}$, $\beta = v/c$, $ \mu= \cos \theta$, $ v$ is the flow velocity,  $\theta$ is the angle formed between the velocity direction and the line of sight  and $c$ is the speed of the light. Moreover, due the bulk motion of the emitting plasma, the radiation received by the Earth's observer is not more isotropic and we must devote our attention in deriving the corrected relation for the transformation of the flux density ($F_\nu$). This work is  attempting to give a further contribution in understanding the physics and origin of GRBs events, incorporating the relativistic motion of the jets and their geometry into the physical models.

\section{The compactness problem}
Currently, an unified model for the jet formation in microquasar, quasar, Blazars, AGNs and GRBs have been proposed. The Blazars with jets pointing directly to us show strong $\gamma$-ray emission. The singular orientation of  jets in these objects conducts us to the fact that we are viewing nearly along the axis of a relativistic outflowing plasma jet; this is  probably the explanation for the enhanced radiation  observed in Blazars.

To explain the ability of high-energy photons to escape the radiating region, the expressions for the absorption optical depth for all photons in $\gamma$-$\gamma$ pair creation process and  $e^- $-$e^+$ pairs of Compton scattering, should be sensitive to the bulk relativistic motion of the radiating moving plasma. The optical depth for both sources of opacity, for high energy values, are of the same order and in the commoving frame a good estimation can be done \citep[][see also Paczynski B. 1986]{pi93,pi99}:

$$\tau' = 2.1\times 10^{11}~ E'^{3/8}_{41} R^{-1/8}_6  exp[{ {-4.5 R^{3/4}_6}\over { E'^{1/4}_{41}}}] \eqno(1)$$

\noindent
where $ E'_{41}$ and  $R_6$  are  the normalized energy and radio of the emitting source.

 In consequence, the optical depths for both absorptive and scattering processes are reduced for the observer by the Doppler factor $\gamma$ as $\tau = \delta ~ \tau' \sim {1\over{2 \gamma^2}}~ \tau'$. Where  $\tau$ is the optical depth in the observer frame and $\tau'$ in the source one; $\delta = {1\over {\gamma (1 -\beta \mu)}}$. This result is in concordance with the expression given by \citet[][see also Piran \& Shemi 1993]{kr91}. Here we have incorporated the observed flatness of the sources assuming a null spectral index for  GRB events ($ F_ \nu \propto \nu ^0$, \citet{ep85}).  
From the Eq. (1) we can see that the medium becomes transparent ($\tau < 1$) for energies $E'< 10^{38}$ ergs.

It is well-accepted that GRBs are extragalactic,  so serious problems with the energetic of these objects have been arisen. To avoid the enormous production of energy, it is assumed that the GRB is collimated. In the cannon-ball model \citep{da02}, it is considered that a whole radiating object moves relativistically with a large factor $\gamma \sim 10^2$ - $10^3$ . Only invoking models with extremely high collimated beams radiating in the direction of the observer, it should be possible to justify the enormous amount of energy coming from a single event. The analysis of GRB collimation was made in  \citet{rho01}.

On the other hand, the relativistic transformation of specific intensity is derived in an elementary way by transforming photon number densities and energy together the relativistic aberration of the angle $\theta$ (angle formed between the line of sight and the flux direction).  Therefore, for a moving source the observed monochromatic flux density $F_\nu$ is related to the flux density in the commoving frame  $F'_\nu$  by  $F_\nu = ({\nu \over{\nu'}})^3  F'_\nu$, where $\nu$ is the observed frequency and $\nu'$ is the frequency related to commoving frame. Consequently, the total fluxes are connected  in both frames by   $F = (\delta^4)  F'$, where $ \delta$ is the Doppler Lorentz factor.

The luminosity $L$, integrating the total flux on a closed surface, gives us the power of energy released by the source. In most of the proposed models the isotropic radiation can not provide the release of energy necessary for the appearance of a cosmological GRB \citep{bi06}. Therefore, we must considerer an anisotropic distribution for the observed radiation. This procedure is performed below.

\section{Anisotropic model}

A lot of questions related to mysterious GRBs  have been raised by  many researchers over two decades. In this context, some explanations have been given; however, we are still far from a complete understanding of the physical nature of these high energetic sources. 

In this part of the current paper, we propose an alternative model derived mainly from relativistic considerations, in order to confront the energetic problem of GRBs, their origin and variability among other consequences.

We star with the assumption that in the rest frame of the source the radiation is released isotropically, then the flux $F'$ does not depend on the angular coordinates (i.e., $F'= F'(r)$).  However, the radiation is emitted from a source which is moving highly relativistically (fireball/relativistic blast wave model, internal shocks), then in concordance with that exposed by others authors \citep[e.g.][]{ghi99,lo04}, the observed radiation must be affected by the Lorentz boosting factor $\delta$.  Therefore, we  detect in the Earth frame a flux $F$ enhanced by the  $\delta$ factor ($ F = \delta^4 ~ F'$), and in the observed frame we should expect an anisotropic flux which is dependent on the propagation direction ($F = F(r,\theta, \phi)$).

The total power emitted by the source and detected by the Earth's observer (luminosity L) can be computed integrating over a closed surface enclosing the source.  The above mentioned fact requires that in this integration we  must incorporate the axial dependence of flux due to the relativistic beaming. Consequently, the integrated power emitted by the source should be expressed as:

$$ L = \oint {\vec F(r,\theta, \phi). \vec{dA} } = \oint {\delta^4 F'  r^2 \sin\theta d\theta d\phi }\eqno(2). $$

\noindent
This gives for the luminosity at the observed frame:

$$ L =  {4 \pi r^2 }~ (4 \gamma^2 - 1)~F'(r)\eqno(3) $$

\noindent 
where  F'(r) is the intrinsic  total flux, which is isotropic. The value of the intrinsic flux $F'(r)$ is determined by the flux registered on the Earth surface, as follow: the GRB observations give us the value of about $10^{-6} {ergs\over{cm^2 sec}}$ - $10^{-7} {ergs\over{cm^2 sec}}$. This flux corresponds to the observed value at $\theta = 0^\circ$ i.e., when the jet orientation coincides with the line of sight (relativistic boosting). Therefore, for $F'(r)$ we will obtain the following expression:

  $$ F'= [\gamma - \sqrt{\gamma ^2 - 1 } ]^4  F(0^\circ),\eqno(4) $$

\noindent
which leads us to deduce, for the observed luminosity, a more appropriate expression:

$$ L =  {4\pi r^2 \over 3} ( 4 \gamma ^2 -1) [\gamma - \sqrt{\gamma ^2 - 1 } ]^4  F(0^\circ).\eqno(5) $$

\noindent
From here, we easily can  obtain the total energy  released  by the source and computed at the observed frame:

$$ E = {4\pi r^2 \over 3} ( 4 \gamma ^2 -1) [\gamma - \sqrt{\gamma ^2 - 1 } ]^4  F(0^\circ) t,\eqno(6) $$

\noindent
where $t$ is  the typical time  GRBs  duration  at the observed frame ($t\sim 1 sec$). 

Finally, the total intrinsic energy of the source at its rest frame can be derived from the previous relation, considering that the integrated energy is also boosted by the  $\delta$ factor as
$E= \delta ^2 E' \sim 2 \gamma E'$, where $E$ is the energy at the observer frame and $ E'$ at the commoving one. Then, the final expression for the intrinsic energy of a GRB event, obtained for  $\gamma \gg 1 $, will be:

$$ E'= {4\pi r^2 \over 3} ( 1 - {1\over {4 \gamma ^2}}) [\gamma - \sqrt{\gamma ^2 - 1 } ]^4  F(0^\circ)~ t. \eqno(7) $$

In the next section we will consider the evaluation of this result and its implications on the GRB physical nature.

\section{Contrains on the GRB energy} 

The expression (7) gives us the true energy that we expect to be released in a GRB event. Inside this relation, the observed flux have been treated taking in consideration the relativistic motion of emitting plasma and its geometry. In order to value this expression, we must take different physical scenarios corresponding to different GRB parameters, which have been published in the literature.

We begin considering that GRBs are located at distances around of $1~ GPc\sim 3\times 10^{27} cm$ and we process computing the energy released by the source for several probable values of observed flux and $\gamma$ Doppler factor . In the same way, we process also with sources located at greater distances, i.e., around $3 ~GPc \sim 1 \times 10^{28} cm $. The obtained results are displayed on the table below.

\begin{deluxetable}{ccrrrl}
\tabletypesize{\scriptsize}
\rotate
\tablecaption{True energy released in a GRB event \label{tbl-1}}
\tablewidth{0pt}
\tablehead{
\colhead{Cosmological Distance (cm)} & \colhead{Observed Flux $F(0^{\circ})$ $(ergs/cm^2~ s)$} & \colhead{Lorentz Factor $\gamma $} & \colhead{ Intrinsic Energy E'$(ergs)$ }}
\startdata
$3 \times 10^{27}$  & $1\times 10^{-6}$ & $ 10$ & $2.37 \times 10^{44}$ &\\
 &  & $100$  & $2.36 \times 10^{40}$ &\\
 &  & $150$  & $4.65\times 10^{39}$ &\\
 &  & $300$  & $2.9\times 10 ^{38}$\\
\hline
$3 \times 10^{27}$  & $1\times  10^{-7}$ & $ 10$ & $2.3 \times 10^{43}$ &\\
  &  & $100$  & $2.36 \times 10^{39}$ &\\
  &  & $150$  & $4.65 \times 10^{38}$ &\\
  &  & $300$  & $2.9\times 10 ^{37}$\\
\hline
$1 \times 10^{28}$  & $1\times 10^{-6}$ & $ 10$ & $2.64 \times 10^{45}$ &\\
  &  & $100$ & $2.62 \times 10^{41}$ &\\
  &  & $150$  & $5.2 \times 10^{40}$ &\\
  &  & $300$  & $3.23\times 10 ^{39}$\\
\hline
$1 \times 10^{28}$  & $1\times 10^{-7}$ & $ 10$ & $ 2.64\times 10^{44}$ &\\
  &  & $100$ & $2.62 \times 10^{40}$ &\\
  &  & $150$  & $5.2 \times 10^{39}$ &\\
  &  & $300$  & $3.23\times 10 ^{38}$\\

\enddata

\tablenotetext{*}{The table contents values for the intrinsic energy of a GRB event at two cosmological distances 
($r\sim 1~ GPc$ and $r\sim 3~ GPc$) and for two observed fluxes ($F (0^\circ) \sim 10^{-6} {ergs\over{cm^2~ sec}}$ and  
$F (0^\circ) \sim 10^{-7} {ergs\over{cm^2~ sec}}$.}
\end{deluxetable}

According with  expression (7) we see that the intrinsic energy of a GRB event is proportional to square of  distance $r$ ($E' \propto r^2$). The same takes place in relation with the observed flux $F(0^\circ)$ ($E'\propto F(0^\circ)$). However, the energy decrees whereas the relativistic moving is increasing ($E'\propto {1\over \gamma}$). From the values presented on the Table 1, we can derive important interpretations on the energetics of GRBs: we note that a GRB can be pretty transparent for mildly cosmological distance ($r\sim 1~GPc$) and for mildly relativistic Lorentz factor ($\gamma \sim 300$). Moreover, for all cases displayed on the Table 1, pleasantly we found that the involved energy in a GRB event is significantly less that the huge energy ($1\times 10^{53}~{ergs\over{cm^2~ sec}}$) required in an isotropic model.

\section{Discussion and Conclusion}
Several authors have made evident the fact that the energy released in a gamma-ray event could be over estimated if the emission is considered isotropic \citep[e.g.][]{fra01,ghi99}. In these models the energy involved is huge, leading to powers of about $10^{52}~{ergs\over{sec}}- 10^{54} ~{ergs\over{sec}}$ in a single event. However, the observations carried out on these explosive events suggest us that anisotropic models are also a good alternative and maybe a more realistic suggestion. In this context, in order to reduce the intrinsic energy involved in a gamma-ray explosion, we may consider gamma-ray burst fireballs having bulk Lorentz factors even larger values than those proposed in literature, i.e., some hundreds or more. However, a Lorentz factor of about hundreds or thousands, being the easy way to diminish the intrinsic energy, it is nothing realistic and we must explain how ejections in GRBs acquire so highly relativistic Lorentz factor. In order to reduce the opacity of the emitting region to pair production and to explain the variability observed in the radiation, a limit to required bulk Lorentz factor $\gamma$ has been established of being be in the range  of 100-300 \citep[e.g.][]{pi99,me00}.

On the other hand, thinking in a relativistic beaming where the observer sees only a limited portion of emitted radiation, we found here a better alternative. Therefore, the true gamma-ray energy must be smaller than the isotropic energy, under consideration that the relativistic beaming makes the observed radiation anisotropic. The radiation in the intrinsic frame can be isotropic, however the observed radiation in the laboratory frame, due to the relativistic beaming, is confined in a very small opening angle and the flux is anisotropic. This second alternative does not need large values for the Lorentz gamma factor and the required intrinsic energy associated with a gamma-ray event is greatly reduced, only considering the anisotropic character of the observed radiation (see above results). Following  this  approach, we derive from an anisotropic radiation point of view, that for relatively small values of $\gamma \sim 10$ the true energy delivered in a gamma-ray event is clustered around $1\times10^{44}~ ergs$, whereas for  $\gamma \sim 100$ the energy is about $1\times10^{40}~ ergs$ and for $\gamma \sim 300$ around $1\times10^{38}~ ergs$, values which are significantly smaller than those commonly presented in the literature.

Our results show that the true energy released in a gamma-ray event should be smaller than the typical electromagnetic and kinetic energy produced of ordinary supernovae. Therefore, the GRB events are not necessarily associated with the formation of stellar black holes, as this was suggested. The values present in this paper are more compatible with the energies  involved in AGN events, where a fraction of a solar mass per year can be accelerated to $\gamma \sim 10$, leading to powers of $\sim 10^{46} ~ {ergs\over {sec}}$.

\end{document}